\documentclass[12pt,twocolumn]{iopart}
\usepackage{amsmath, amssymb, amsxtra, mathrsfs}
\usepackage{graphics, epsfig, here, fancyhdr,graphicx}
\usepackage{bbm, color}
\usepackage[all]{xy}
\usepackage{subfigure}
\usepackage{setspace}
\usepackage{enumerate}

\newcommand{\bN}{\mathbb{N}}

\newcommand{\bR}{\mathbb{R}}

\newcommand{\cP}{\mathcal{P}}

\newcommand{\cJ}{\mathcal{J}}

\newcommand{\cD}{\mathcal{D}}

\newcommand{\cO}{\mathcal{O}}

\newcommand{\ol}[1]{{\overline{#1}}}
\newcommand{\dS}{\displaystyle}

\newcommand{\bc}[1]{{\left\{#1\right\}}}
\newcommand{\bq}[1]{{\left[#1\right]}}
\newcommand{\bp}[1]{{\left(#1\right)}}

\newcommand{\bt}[1]{{\left\langle#1\right\rangle}}

\newcommand{\set}[2]{{\bc{#1\,\,:\,\,#2}}}

\newcommand{\teQ}{\triangleq}

\newcommand{\rM}[1]{{\mathrm{#1}}}

\newcommand{\of}{\subseteq}

\newcommand{\vphi}{\varphi}
\newcommand{\Dx}{\dot{x}}

\renewcommand{\P}{\mathbb{P}}

\newcommand{\w}{\omega}
\newcommand{\cF}{\mathcal{F}}

\newcommand{\st}{\,\,\mathrm{ s.t. }\,\,}

\newcommand{\eP}{\varepsilon}

\newcommand{\tw}[1]{{\widetilde{#1}}}
\newcommand{\wh}[1]{{\widehat{#1}}}

\newcommand{\half}{\frac{1}{2}}
\newcommand{\tH}{^{\mathrm{th}}}

\newcommand{\rd}{\mathrm{d}}

\newcommand{\dxdy}[2]{{\frac{\rd #1}{\rd #2}}}

\begin{document}

\title[Optimal Entrainment of Neural Oscillator Ensembles]{Optimal Entrainment of Neural Oscillator Ensembles}

\author{Anatoly Zlotnik, Jr-Shin Li}


\address{Department of Electrical and Systems Engineering\\
    Washington University\\
    Saint Louis, Missouri 63130\\}
\ead{azlotnik@ese.wustl.edu, jsli@ese.wustl.edu}
\begin{abstract}
In this paper, we derive the minimum-energy periodic control that entrains an ensemble of structurally similar neural oscillators to a desired frequency.  The state space representation of a nominal oscillator is reduced to a phase model by computing its limit cycle and phase response curve, from which the optimal control is derived by using formal averaging and the calculus of variations.  We focus on the case of a 1:1 entrainment ratio, and introduce a numerical method for approximating the optimal controls.  The method is applied to asymptotically control the spiking frequency of neural oscillators modeled using the Hodgkin-Huxley equations.  This illustrates the optimality of entrainment controls derived using phase models when applied to the original state space system, which is a crucial requirement for using phase models in control synthesis for practical applications.  The results of this work can be used to design low energy signals for deep brain stimulation therapies for neuropathologies, and can be generalized for optimal frequency control of large-scale complex oscillating systems with parameter uncertainty.
\end{abstract}

\textfloatsep=20pt
\submitto{\JNE}
\maketitle

\section{Introduction} \label{secintro}

The synchronization of oscillating systems is an important
and extensively studied scientific concept with
numerous engineering applications \cite{strogatz01}.  Examples
include the oscillation of neurons \cite{hoppensteadt97}, sleep
cycles and other pacemakers in biology
\cite{hanson78,mirollo90,ermentrout84}, semiconductor
lasers in physics \cite{fischer00}, and vibrating systems in
mechanical engineering \cite{blekhman88}.  Among many well-studied synchronization phenomena, the asymptotic synchronization of an oscillator to a periodic control signal,
called entrainment, is of fundamental scientific and engineering importance \cite{aronson86,zalalutdinov03}.  The intrinsic occurrence and extrinsic imposition of entrainment in networked oscillators is of particular interest in neuroscience \cite{berke04,sirota08}.  It has been investigated as an important mechanism underlying the coordinated resetting of neural subpopulations, which is required for demand-controlled deep brain stimulation (DBS) for clinical treatment of Parkinson's disease \cite{tass03}.  In clinical DBS, entrainment is accomplished using a sequence of pulses with tunable amplitude, duration, and frequency, after which another effectively desynchronizing stimulation is applied. Because low energy consumption increases implant battery life, decreases tissue damage, and minimizes side effects and risks \cite{kuncel04}, the development of minimum energy stimuli for the entrainment of neural oscillator populations is critical to improving the clinical outcomes of DBS.

The entrainment, and hence frequency control, of an oscillating system can be examined by considering its phase response curve (PRC) \cite{izhikevich06,izhikevich07}, which
quantifies the shift in asymptotic phase due to an infinitesimal
perturbation in the state.  
The classic phase coordinate transformation \cite{malkin49} for studying nonlinear oscillators
was used together with formal averaging \cite{kornfeld82} to develop
a model of synchronization in coupled chemical oscillations \cite{kuramoto84}.  Since then, phase models have become indispensable in physics, chemistry, and biology
 for studying oscillating systems where the full state-space model is complicated or even unknown, but where the phase 
can be estimated from partial state observations, and the PRC can be approximated experimentally \cite{pikovsky01}.  They have been successfully applied to investigate many synchronization phenomena \cite{strogatz00}, focusing on synchronization emerging in networks of interacting oscillators and on the response of large collections of oscillators to periodic external stimuli \cite{rosenblum96,hong02}.  Such models have long been of interest to neuroscientists \cite{pinsker77,ermentrout96}, who have been motivated by the prospect of using dynamical systems theory to improve the effectiveness of DBS as a clinical therapy for epilepsy and Parkinson's disease \cite{perlmutter06,good09}.  Several studies have concentrated on the use of phase models in order to attain desired design objectives for electrochemical \cite{kiss02,nakata09} and neural \cite{hoppensteadt99,hunter03} systems, including recent work that approaches the use of phase models in neuroscience from a control theoretic perspective \cite{ullah09,nabi09}.  The control of neural spiking using minimum energy controls with constrained amplitude and charge balancing has also recently examined \cite{dasanayake11,dasanayake12}.  These studies have demonstrated that phase-model reduction provides a practical approach to synthesizing near-optimal controls that achieve design goals for oscillating neural systems.

Much of the work on the control of neural oscillators is based on the assumption that each neuron behaves according to pre-defined underlying dynamics, such as the Hodgkin-Huxley equations \cite{hodgkin52}, which constitute a widely studied model of action potential propagation in a squid giant axon.  However, in practical applications of neural control and engineering, the systems in question are collections of biological neurons that exhibit variation in parameters that characterize the system dynamics, specifically the frequency of oscillation and sensitivity to external stimuli.  Although such a system consists of a finite collection of subsystems, it contains so many unobservable elements, each with parameter uncertainty, that its collective dynamics are most practically modeled by indexing the subsystems by a parameter varying on a continuum.  The control of such neural systems therefore lies within an emerging and challenging area in mathematical control theory called ensemble control, which encompasses a class of problems involving the guidance of an uncountably infinite collection of structurally identical dynamical systems with parameter variation by applying a common open-loop control \cite{li06thesis}.  In the context of phase model reduction, the appropriate indexing parameter for such a collection of oscillating systems is natural frequency.  Therefore, a practical approach to the optimal design of inputs that entrain a collection of neurons is to first consider a family of phase models with common nominal PRC and frequency varying over a specified interval.   Optimal waveforms that entrain a collection of phase oscillators with the greatest range of frequencies by weak periodic forcing have been characterized for certain oscillating chemical systems \cite{harada10}, and this approach has been extended to a method for optimal entrainment of oscillating systems with arbitrary PRC \cite{zlotnik11}.

In this paper, we
develop a method for engineering weak, periodic signals that entrain ensembles of structurally similar uncoupled oscillators with variation in system parameters to a desired target frequency without the use of state feedback.  
In addition, we present an efficient numerical method for approximating optimal waveforms by minimizing over a compact interval a polynomial whose coefficients depend on the PRC of the entrained oscillator.  
A related computation is performed to approximate the region in the energy-frequency plane in which entrainment of an oscillator with a given PRC by a particular waveform occurs. Such graphs, called Arnold tongues \cite{pikovsky01}, are used to characterize the performance of controls derived using the PRC for entrainment of the original oscillator in state space, which is the ultimate purpose of using phase-model based control.  We conduct this important validation, which is largely lacking in the literature, using the Hodgkin-Huxley equations as an example.  The results of this work can also be viewed as a method for constructing a control to optimally shape the Arnold tongue for an entrainment task involving an ensemble of neurons.

In the following section, we discuss the phase coordinate transformation for a nonlinear oscillator and the available numerical methods for computing the PRC.  In Section \ref{secent}, we describe how averaging theory is used to study the asymptotic behavior of an oscillating system, and use the calculus of variations to derive the minimum energy entrainment control for a single oscillator with arbitrary PRC.  In Section \ref{secens}, we formulate and solve the problem of minimum energy entrainment of oscillator ensembles, for which the optimal controls can be synthesized by using an efficient procedure involving Fourier series and Chebyshev polynomials detailed in Appendix A.  Throughout the paper, important concepts are illustrated graphically by providing examples using the Hodgkin-Huxley model, which is described in Appendix B, and its corresponding PRC.  We provide computed Arnold tongues in addition to those derived using our theory, which verify that the optimal inputs derived here are a significant improvement on commonly used entrainment waveforms.  In Section \ref{secneur}, 
we describe several computational results that provide further justification for our approach.  Finally in Section \ref{secconc}, we discuss our conclusions and future extensions of this work.   

\section{Phase models} \label{secpv}

The phase coordinate transformation is a well-studied model reduction technique that is useful for studying oscillating systems characterized by complex nonlinear dynamics, and can also be used for system identification when the dynamics are unknown.  Consider a full state-space model of an oscillating system, described by a smooth ordinary differential equation system
\begin{equation} \label{sys1}
\Dx=f(x,u), \quad x(0)=x_0,
\end{equation}
where $x(t)\in\bR^n$ is the state and $u(t)\in\bR$ is a control.  Furthermore, we require that (\ref{sys1}) has an attractive, non-constant limit cycle $\gamma(t)=\gamma(t+T)$, satisfying $\dot{\gamma}=f(\gamma,0)$, on the periodic orbit $\Gamma=\set{y\in\bR^n}{y=\gamma(t) \text{ for } 0\leq t< T}\subset\bR^n$.    In order to study the behavior of this system, we reduce it to a scalar equation
\begin{equation} \label{sys2}
\dot{\psi}=\w+Z(\psi)u,
\end{equation}
which is called a phase model, where $Z$ is the phase response curve (PRC) and $\psi(t)$ is the phase associated to the isochron on which $x(t)$ is located. The isochron is the manifold in $\bR^n$ on which all points have asymptotic phase $\psi(t)$ \cite{brown04}.  The conditions for validity and accuracy of this model have been determined \cite{efimov10}, and the reduction is accomplished through the well-studied process of phase coordinate transformation \cite{efimov09}, which is based on Floquet theory \cite{perko90,kelley04}.  The model is assumed valid for inputs $u(t)$ such that the solution $x(t,x_0,u)$ to (\ref{sys1}) remains within a neighborhood $U$ of $\Gamma$.

\begin{figure}[t]
\centerline { \includegraphics[width=.6\linewidth]{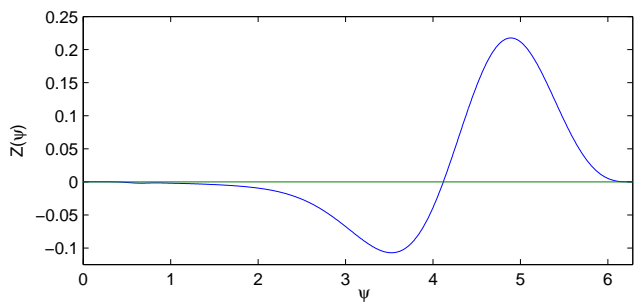}
} \caption{Hodgekin-Huxley phase response curve (PRC). The natural period and frequency of oscillation are $T\approx 14.638$ and $\w\approx0.429$, respectively.} \label{figprc1}
\end{figure}

To compute the PRC, the period $T=2\pi/\w$ and the limit cycle $\gamma(t)$ must be approximated to a high degree of accuracy.  This can be done using a method for determining the steady-state response of nonlinear oscillators \cite{aprille72} based on perturbation theory \cite{khalil02} and gradient optimization \cite{peressini00}.
The PRC can then be computed by integrating the adjoint of the linearization of (\ref{sys1}) \cite{ermentrout96}, or by using a more efficient and numerically stable spectral method developed more recently \cite{govaerts06}.  A software package called XPPAUT \cite{ermentrout02} is commonly used by researchers to compute the PRC.  The PRC of the Hodgkin-Huxley system with nominal parameters, obtained using a technique derived from the method of Malkin \cite{malkin49}, is displayed in Figure \ref{figprc1}.

\section{Entrainment of oscillators} \label{secent}

Suppose that one desires to entrain the system (\ref{sys2}) to a new frequency $\Omega$ using a periodic control $u(t)=v(\Omega t)$ where $v$ is $2\pi$-periodic.   We have adopted the weak forcing assumption, i.e., $v=\eP v_1$ where $v_1$ has unit energy and $\eP<<1$, so that given this control the state of the original system (\ref{sys1}) is guaranteed to remain in a neighborhood $U$ of $\Gamma$ in which the phase model (\ref{sys2}) remains valid \cite{efimov10}.  Now define a slow phase variable by $\phi(t)=\psi(t)-\Omega t$, and call the difference $\Delta\w=\w-\Omega$ between the natural and forcing frequencies the frequency detuning.  The dynamic equation for the slow phase is
\begin{equation} \label{sys3}
\dot{\phi}=\dot{\psi}-\Omega=\Delta\w+Z(\Omega t+\phi)v(\Omega t),
\end{equation}
where $\dot{\phi}$ is called the phase drift.  In order to study the asymptotic behavior of (\ref{sys3}) it is necessary to eliminate the explicit dependence on time on the right hand side, which can be accomplished by using formal averaging \cite{kuramoto84}.  Given a periodic forcing with frequency $\Omega=2\pi/T$, we denote the forcing phase $\theta=\Omega t$.   If $\cP$ is the set of $2\pi$-periodic functions on $\bR$, we can define an averaging operator $\bt{\cdot}:\cP\to\bR$ by
\begin{equation} \label{ave}
\bt{x}=\frac{1}{2\pi}\int_0^{2\pi}x(\theta)\rd \theta.
\end{equation}
The weak ergodic theorem for measure-preserving dynamical systems on the torus \cite{kornfeld82} implies that for any periodic function $v$, the interaction function
\begin{align} \label{Lambdef}
\Lambda_v(\phi) & = \bt{Z(\theta+\phi)v(\theta)} \notag\\ & =
\frac{1}{2\pi}\int_0^{2\pi} Z(\theta+\phi)v(\theta)\rd \theta \\ & =\lim_{T\to\infty}\frac{1}{T}\int_0^TZ(\Omega t+\phi)v(\Omega t)\rd t \notag
\end{align}
exists as a smooth, $2\pi$-periodic function in $\cP$.  By the formal averaging theorem \cite{hoppensteadt97}, the system
\begin{equation} \label{sys4}
\dot{\vphi}=\Delta \w + \Lambda_v(\vphi)+\cO(\eP^2)
\end{equation}
approximates (\ref{sys3}) in the sense that there exists a change of variables $\vphi=\phi+\eP h(\vphi,\phi)$ that maps solutions of (\ref{sys3}) to those of (\ref{sys4}).  Therefore the weak forcing assumption $v=\eP v_1$ with $\eP<<1$ allows us to approximate the phase drift equation by
\begin{equation} \label{sys5}
\dot{\vphi}=\Delta \w + \Lambda_v(\vphi).
\end{equation}

\begin{figure}[t]
\centerline { \includegraphics[width=.6\linewidth]{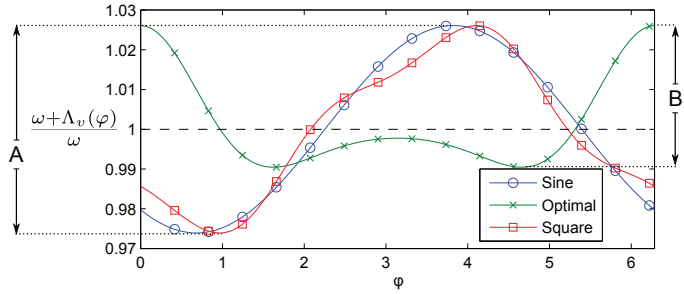}}
\caption{Translated, rescaled interaction functions of the Hodgkin-Huxley PRC with sinusoidal ($v_\circ$) and square wave ($v_{_\square}$) inputs of RMS energy $P_{RMS}=0.2$, and with the optimal waveform $v_+$ for increasing frequency ($\times$) of energy $P_{RMS}=0.1301$.  The range of target frequencies $\Omega$ to which $v_\circ$ and $v_+$ can entrain the system are indicated by $A$ and $B$, respectively.  Equation (\ref{sys5}) has a fixed point for values of $\Omega$ up to $2.6\%$ greater than $\w$ for each control, but the optimal control requires $35\%$ less RMS energy.} \label{figinter1}
\end{figure}

The averaged equation (\ref{sys5}) is independent of time, and can be used to study the asymptotic behavior of the system (\ref{sys2}) under periodic forcing.  We say that the system is entrained by a control $u=v(\Omega t)$ when the phase drift equation (\ref{sys5}) satisfies $\dot{\vphi}=0$.  This will eventually occur if there exists a phase $\vphi_*$ satisfying $\Delta \w + \Lambda_v(\vphi_*) = 0$.  The range of frequencies to which the Hodgkin-Huxley system can be entrained by several waveforms is illustrated in Figure \ref{figinter1}.  Because $\Lambda_v(\vphi)$ is nontrivial, when the system is entrained there exists at least one $\vphi_*\in[0,2\pi)$ that is an attractive fixed point of (\ref{sys5}).  In practical applications, it is desirable to achieve entrainment with a control of minimum energy.   By defining the phases $\vphi_-=\arg\min_\vphi \Lambda_v(\vphi)$ and $\vphi_+=\arg\max_\vphi\Lambda_v(\vphi)$, we can formulate minimum energy entrainment of an oscillator as a variational optimization problem.  The objective function to be minimized is the energy $\bt{v^2}$, and entrainment can be achieved when $\w+\Lambda_v(\vphi_+)\geq \Omega$ if $\Omega>\w$ and $\w+\Lambda_v(\vphi_-)\leq\Omega$ if $\Omega<\w$.  This inequality is active for the optimal waveform, and hence can be expressed as the equality constraint
\begin{equation} \label{const1}
\begin{array}{rcl}
\Delta \w+\Lambda_v(\vphi_+)= 0  &\quad \text{if}\quad & \Omega>\w,\\\Delta \w+\Lambda_v(\vphi_-)= 0 &\quad\text{if}\quad& \Omega<\w.
\end{array}
\end{equation}
We formulate the problem for $\Omega>\w$ to obtain the minimum energy control $v_+$ using the calculus of variations \cite{gelfand00}.  The derivation of the case where $\Omega<\w$ is similar, and results in the symmetric control $v_-$. The constraint (\ref{const1}) can be adjoined to the objective using a multiplier $\lambda$, resulting in the cost\\
\begin{eqnarray} \label{op1}
\cJ[v] & = & \bt{v^2} - \lambda(\Delta\w + \Lambda_v(\vphi_+)) \\ & = &
\bt{v^2} - \lambda\bp{\Delta\w + \frac{1}{2\pi} \int_0^{2\pi} Z(\theta+\vphi_+)v(\theta)\rd \theta} \notag\\
& = & \frac{1}{2\pi} \int_0^{2\pi} \bq{v(\theta)(v(\theta)-\lambda Z(\theta+\vphi_+)) - \lambda\Delta \w}\rd \theta. \notag
\end{eqnarray}\\
Applying the Euler-Lagrange equation, we obtain the necessary condition for an optimal solution, which yields a candidate function
\begin{equation} \label{sol0}
v(\theta)=\frac{\lambda}{2} Z(\theta+\vphi_+),
\end{equation}
which we substitute into the constraint (\ref{const1}) and solve for the multiplier, $\lambda=-2\Delta \w/\bt{Z^2}$.  Consequently the minimum energy controls are
\begin{equation} \label{sol1}
\begin{array}{ll} v_+(\theta) = \dS-\frac{\Delta \w}{\bt{Z^2}} Z(\theta+\vphi_+) &  \quad \text{if}\quad  \Omega>\w,\\ \\\dS v_-(\theta) =-\frac{\Delta \w}{\bt{Z^2}} Z(\theta+\vphi_-) &  \quad \text{if}\quad  \Omega<\w. \end{array}
\end{equation}
In practice we omit the phase ambiguity $\vphi_+$ or $\vphi_-$ in the solution (\ref{sol1}) because entrainment is asymptotic.

In addition to deriving the optimal control, we are interested in viewing the Arnold tongue, which is a plot of the minimum root mean square (RMS) energy $P_v(\Omega)=\sqrt{\bt{v^2}}$ required for entrainment of the system by a control $v$ to a given target frequency $\Omega$.  This is accomplished by substituting into (\ref{const1}) the expression $v(\theta)=P_v(\Omega)v_1(\theta)$, where $v_1$ is a unity energy normalization of $v$.  This results in
\begin{equation}
\begin{array}{rcl}
\Delta \w+\Lambda_{v_1}(\vphi_+)P_v(\Omega)= 0  &\quad \text{if}\quad & \Omega>\w,\\\Delta \w+\Lambda_{v_1}(\vphi_-)P_v(\Omega)= 0 &\quad\text{if}\quad& \Omega<\w,
\end{array}
\end{equation}
which in turn yields the Arnold tongue boundary estimate given by
\begin{equation}
P_v(\Omega)=\left\{\begin{array}{lll} -\Delta\w/\Lambda_{v_1}(\vphi_+) & \quad \text{if}\quad & \Omega>\w \\ -\Delta\w/\Lambda_{v_1}(\vphi_-) &\quad\text{if}\quad& \Omega<\w. \end{array}
\right.
\end{equation}

\begin{figure}[t]
\centering \subfigure[]{\includegraphics[width=.6\linewidth]{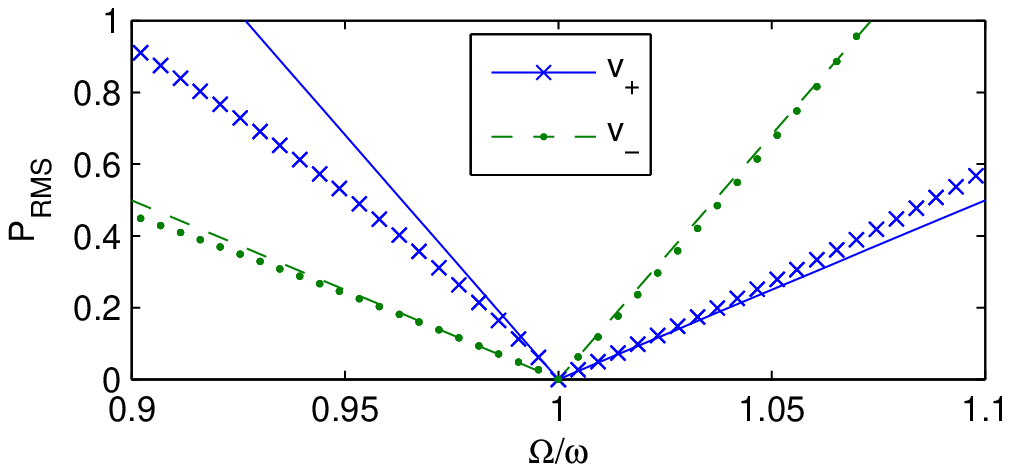} \label{atong2a}} \\ \subfigure[]{\includegraphics[width=.6\linewidth]{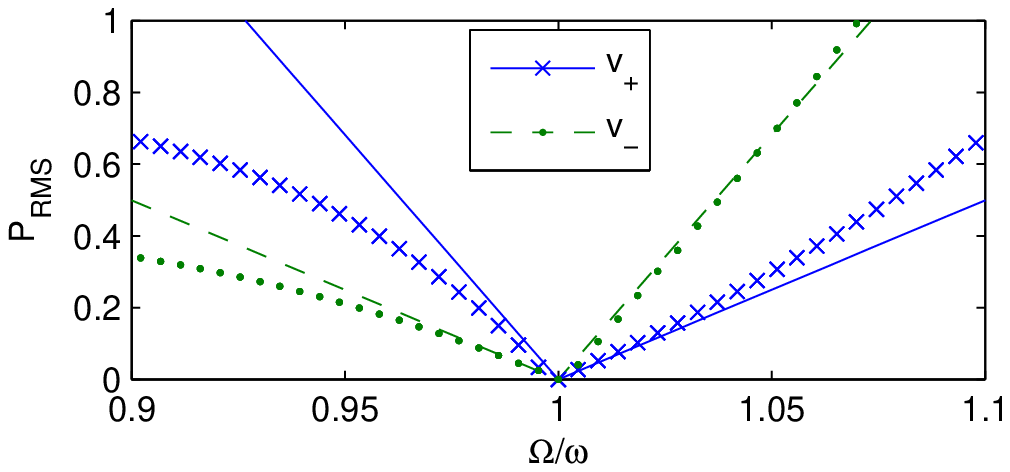} \label{atong2b}}
 \caption{Arnold tongues for optimal waveforms for frequency increase ($v_+$) and decrease ($v_-$) of the Hodgkin-Huxley oscillator.  Theoretical boundaries predicted by phase reduction theory are shown as lines, and values computed using \ref{atong2a} the Hodgkin-Huxley PRC and \ref{atong2b} the Hodgkin-Huxley equations are shown as points.  These values are computed using a line search over the RMS forcing energy, and \ref{atong2a} closely approximates \ref{atong2b}.} \label{figatong1}
\end{figure}

The boundaries of the theoretical Arnold tongues for the controls $v_+$ and $v_-$, as well as computed values of the RMS forcing energy required to entrain the Hodgkin-Huxley system using these controls, are shown in Figure \ref{figatong1}.

In this section, we have shown that the minimum energy periodic control $u(t)=v(\theta)$ that entrains a single oscillator with natural frequency $\w$ to a target frequency $\Omega$ is a re-scaling of the PRC, where $\theta=\Omega t$ is the forcing phase.  Observe that this control will entrain oscillators with natural frequencies between $\w$ and $\Omega$ to the target $\Omega$ as well.  In the following section, we derive the minimum energy periodic control that entrains each member of a family of oscillators, all of which share the same phase response curve but which can have a natural frequency taking any value on a specified interval, to a single target frequency $\Omega$.

\section{Entrainment of an ensemble of oscillators} \label{secens}

We now extend the above approach to derive a single minimum energy periodic control signal $v(\Omega t)$ that guarantees entrainment for each system in the ensemble of oscillators
\begin{equation}
\cF=\{\dot{\psi}=\w+Z(\psi)u\,\,:\,\,\w\in(\w_1,\w_2)\}
\end{equation}
to a frequency $\Omega$.  We call the range of frequencies that are entrained by the control $v$ the locking range $R[v]=[\w_-,\w_+]$, and when $(\w_1,\w_2)\subset R[v_*]$ we say that the ensemble $\cF$ is entrained.   This requirement results in the constraints
\begin{equation} \label{const2}
\begin{array}{rcrcrcrcr}
\Delta \w_+ &\teQ & \w_+-\Omega &=&-\Lambda_v(\vphi_-) &\geq& \w_2-\Omega &\teQ & \Delta \w_2,  \\
\Delta \w_- &\teQ &\w_--\Omega &=&-\Lambda_v(\vphi_+) & \leq &  \w_1-\Omega & \teQ & \Delta \w_1.
\end{array}
\end{equation}

\begin{figure}[t]
\centering \subfigure[]{\includegraphics[width=.6\linewidth]{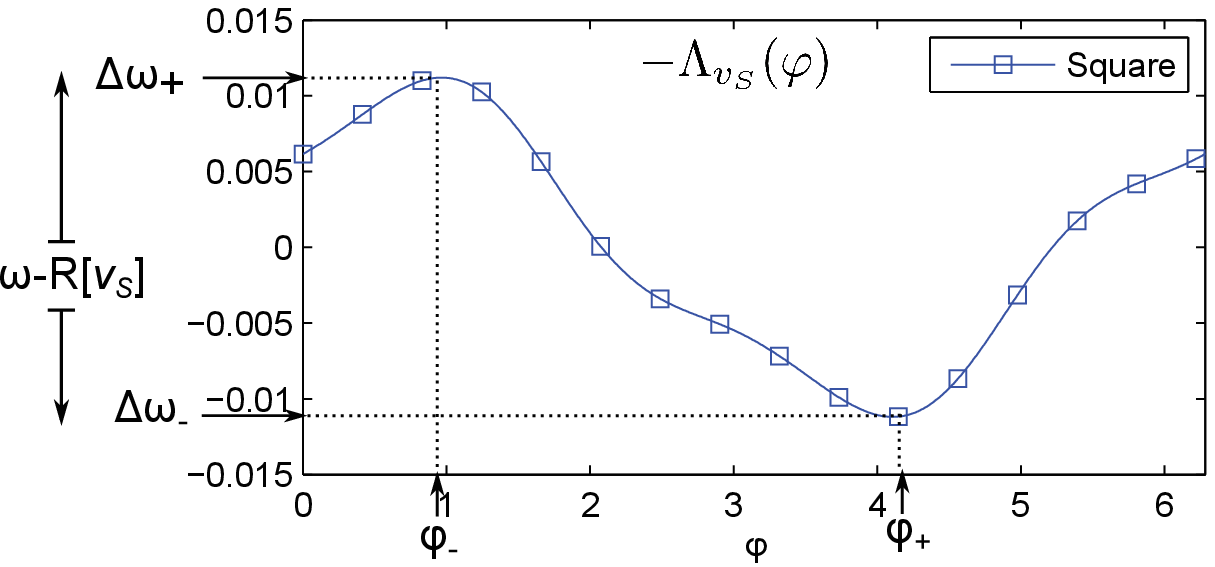} \label{detun2a}} \\ \subfigure[]{\includegraphics[width=.6\linewidth]{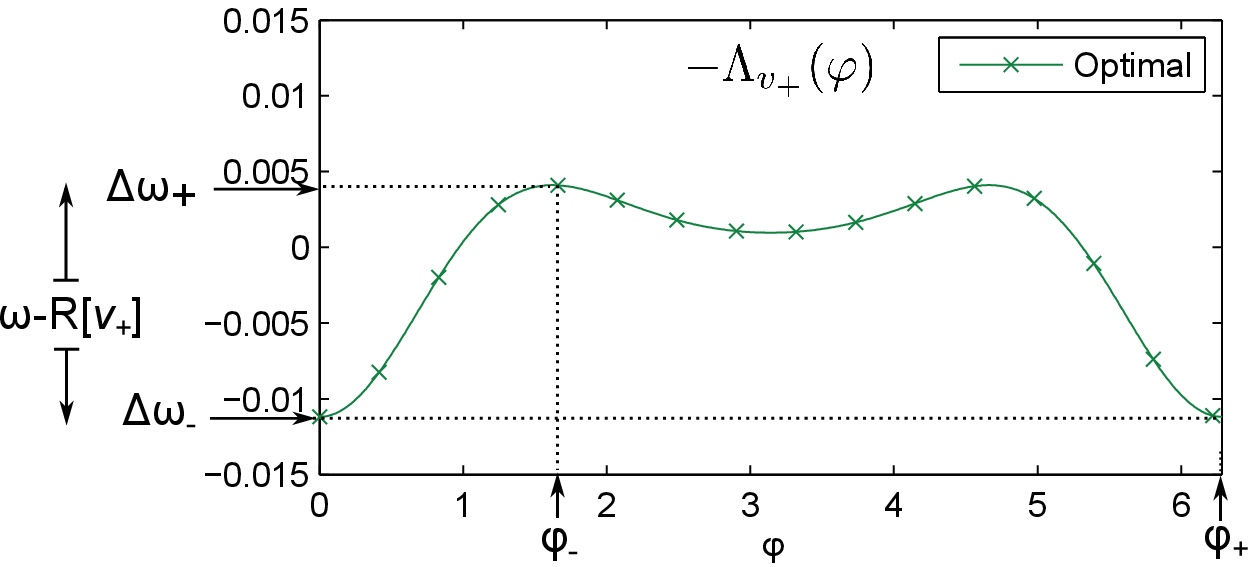} \label{detun2b}} \caption{The negative of the interaction function ($-\Lambda_v$) of the Hodgkin-Huxley PRC \ref{detun2a} with a square wave ($\square$) input $v_S$ of RMS energy $P=0.2$ and \ref{detun2b} with the optimal waveform $v_+$ for increasing frequency ($\times$) of energy $P=0.1301$.  The control $v_S$ achieves a locking range $R[v_S]=[\w-0.0112,\w+0.0112]$ that is symmetric about the natural frequency $\w$, while the control $v_*$ achieves a non-symmetric locking range $R[v_*]=[\w-0.0041,\w+0.0112]$.  For target frequencies $\Omega$ near $\w_2$, $v_+$ can entrain $\cF$ using 35\% less energy.  We require $\Delta\w_2<\Delta\w_+$ and $\Delta\w_1>\Delta\w_-$ to guarantee entrainment of the ensemble.} \label{figdetun1}
\end{figure}

Note that $\Omega\in[\w_-,\w_+]$ always holds, because an oscillator with natural frequency $\Omega$ is always entrained when forced with the same frequency.  The relationship between the interaction function $\Lambda_v$, the locking range $R[v]$, and the frequency dispersion $(\w_1,\w_2)$ of the family $\cF$ is illustrated in Figure \ref{figdetun1}.  The objective of minimizing control energy $\bt{v^2}$ given the constraints (\ref{const2}) gives rise to the optimization problem
\begin{equation} \label{op2}
\begin{array}{rl}
\min & \cJ[k] = \bt{v^2}, \quad k\in\cP\medskip\\
\st &\,\,\,\, \Delta \w_2+\Lambda_v(\vphi_-) \leq 0, \medskip \\
 & -\Delta \w_1-\Lambda_v(\vphi_+) \leq 0.
\end{array}
\end{equation}
If $\w_2<\Omega$ (resp. $\w_1>\Omega$), then problem (\ref{op2}) is solved by the control $v_+$ where $\Delta\w=\Delta\w_1$ (resp. $v_-$ where $\Delta\w=\Delta\w_2$) in (\ref{sol1}), however these are not the only instances where that solution is optimal.  See, for instance, case (B) in Figure \ref{arnold3a}.  Understanding the Arnold tongue that characterizes the entrainment of the ensemble $\cF$ by a control $v$ will clarify the condition when (\ref{sol1}) is optimal.  We derive this condition, which depends on the ensemble parameters $\w_1$ and $\w_2$ as well as the target frequency $\Omega$, and then consider the case in which another class of optimal solutions is superior.  These two cases are illustrated in Figure \ref{arnold3b}.

\subsubsection{Case I: A re-scaled PRC is optimal for entrainment of the ensemble.}

To derive the conditions when (\ref{sol1}) is optimal, we focus on the use of $v_-$  to entrain $\cF$ to a frequency $\Omega\in(\w_1,\w_2)$ when $\Delta\w_+=\Delta\w_2>-\Delta\w_1$, noting that the case where $\Delta\w_2<-\Delta\w_1=-\Delta\w_-$ and $v_+$ is used is symmetric.  Because $\w_2$ is the  natural frequency in the ensemble farthest from $\Omega$, we use $\Delta\w=\Delta\w_2$.  Then the first constraint in (\ref{op2}) is active, yielding $-\Delta\w_2=\Lambda_v(\vphi_-)=\Omega-\w_+$, so that $\w_+=\w_2$ is the upper bound on the locking range $R[k]$, as desired. It remains to determine $\Lambda_v(\vphi_+)=\Omega-\w_-$, from which we obtain the lower bound $\w_-$ on $R[k]$.  Let us denote $\Delta \vphi=\vphi_+-\vphi_-$, and define
\begin{equation} \label{qdef}
Q(\Delta\vphi)=\bt{Z(\theta+\Delta\vphi)Z(\theta)}.
\end{equation}
We can define an inner product $(\cdot,\cdot):\cP\times\cP\to\bR$ by $(f,g)=\bt{fg}$, so that the Cauchy-Schwartz inequality yields $|Q(\Delta\vphi)|\leq\bt{Z^2}=Q(0)$.  Furthermore, the periodicity of $Z$ results in $Q(\Delta\vphi)=\bt{Z(\theta+\Delta\vphi)Z(\theta)} = \bt{Z(\theta)Z(\theta-\Delta\vphi)} = Q(-\Delta\vphi)$.   Combining (\ref{Lambdef}), (\ref{sol1}), and (\ref{qdef}), we can write
\begin{equation}
\Lambda_v(\vphi)  =  \bt{Z(\theta+\vphi)v_-(\theta)} =
-\frac{\Delta\w_2}{\bt{Z^2}}Q(\vphi-\vphi_-),
\end{equation}
resulting in
\begin{equation} \label{vminent1}
-\Delta\w_+=\Lambda_v(\vphi_-)=-\Delta\w_2,
\end{equation}
as expected.  Observe that $\Lambda_v(\vphi)$ is maximized when $Q(\vphi-\vphi_-)$ is minimized, and hence to find $\Lambda_v(\vphi_+)$ it suffices to find the minimum value $Q_*$ of $Q(\Delta\vphi)$.  Assume for now that $Q_*<0$, which is true for the Hodgkin-Huxley PRC, and typical for Type II neurons.  The practical considerations for finding $Q_*$ are discussed in appendix B.  It follows that
\begin{equation} \label{bound1}
\Lambda_v(\vphi_+)=-\frac{\Delta\w_2}{\bt{Z^2}}Q_*,
\end{equation}
and the lower bound of $R[k]$ is $\w_-=\Omega-\Lambda_v(\vphi_+)$. If $\w_-<\w_1$, then $(w_1,w_2)\subset R[k]$, hence the control $v_-$ in (\ref{sol1}), with $\Delta\w = \w_2-\Omega$, is the minimum energy solution to problem (\ref{op2}), and entrains $\cF$ to the frequency $\Omega$.

Now let us define $v(\theta)=P_{v}(\w)\tw{v}(\theta)$ where $\tw{v}$ is the unity energy normalization of a control $v$.  Substituting this expression into (\ref{const2}), we can obtain the minimum RMS energy $P_{v_-}(\w)$ required to entrain the member of $\cF$ with a natural frequency $\w$ to $\Omega$ using the control $v_-$.  Because $\Lambda_v(\vphi)=\Lambda_{\tw{v}}(\vphi)P_v(\w)$, this yields
\begin{equation}\label{ensat1}
\begin{array}{rcl}
\Delta \w_++\Lambda_{\tw{v}_-}(\vphi_-)P_{v_-}(\w) & = 0, & \w > \Omega,\\
\Delta\w_-+\Lambda_{\tw{v}_-}(\vphi_+)P_{v_-}(\w) & = 0, & \w < \Omega.
\end{array}
\end{equation}
Because $\tw{v}_-=Z/\sqrt{\bt{Z^2}}$, it follows that $\Lambda_{\tw{v}_-}(\vphi)\sqrt{\bt{Z^2}}=Q(\vphi-\vphi_-)$,
and substituting this into (\ref{ensat1}) and solving for $P_{v_-}(\w)$ yields
\begin{equation}
P_{v_-}(\w)= \left\{\begin{array}{cl} \dfrac{1}{\sqrt{\bt{Z^2}}}(\w-\Omega) &  \quad \text{if}\quad  \Omega<\w,\\ \\\dfrac{\sqrt{\bt{Z^2}}}{Q_*}(\w-\Omega) &  \quad \text{if}\quad  \Omega>\w. \end{array}\right.
\end{equation}

\begin{figure}[t]
\centering \subfigure[]{\includegraphics[width=.6\linewidth]{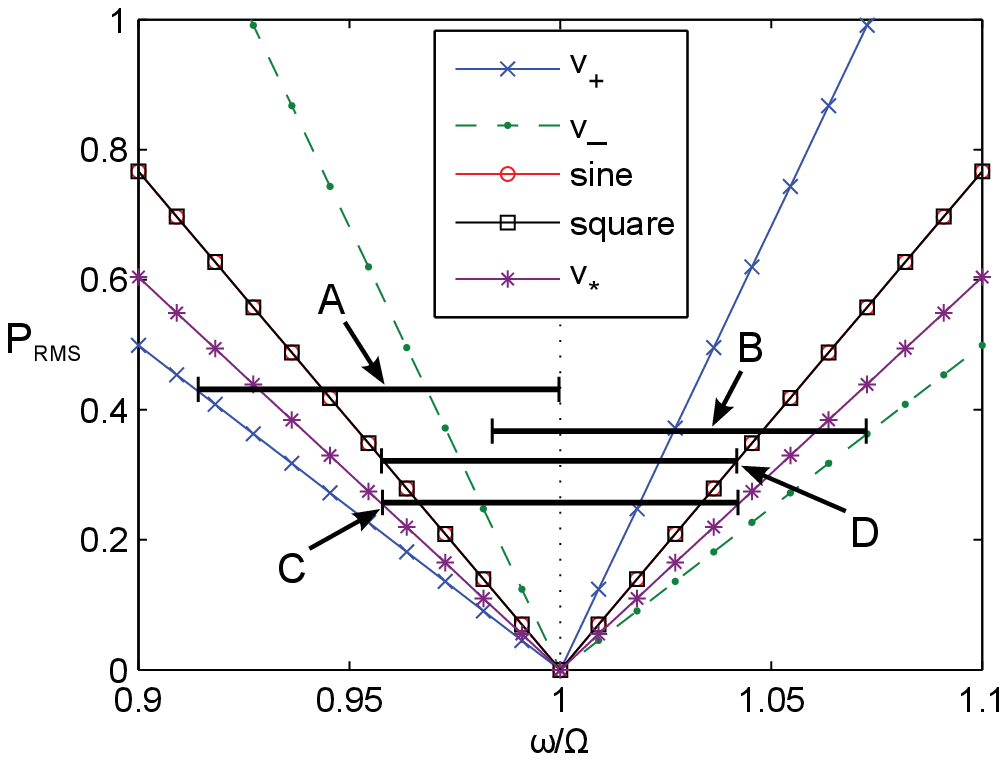} \label{arnold3a}} \\ \subfigure[]{\includegraphics[width=.6\linewidth]{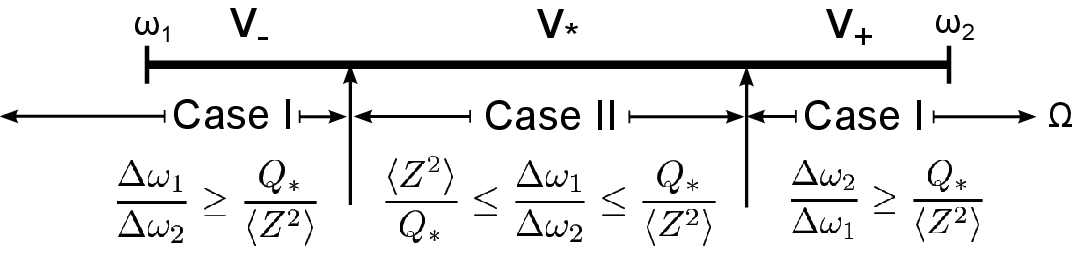} \label{arnold3b}} \caption{\ref{arnold3a} Arnold tongues of Hodgkin-Huxley neuron ensembles for several controls, where the target $\Omega$ is the nominal natural frequency.  The span of each horizontal bar represents the range of natural frequencies of a family $\cF$ of oscillators with the Hodgkin-Huxley PRC.  The vertical location of each bar represents the minimum RMS energy required to entrain $\cF$ by the indicated waveform.  The control $v_+$ is optimal for ensemble (A), and $v_-$ is optimal for ensemble (B).  The waveform $v_*$, which is optimal for entraining the family (C) to $\Omega=\half(\w_1+\w_2)$, achieves entrainment with RMS energy 0.26, which is 21\% lower than the RMS energy 0.33 required to do so with a sine or square wave (D). \ref{arnold3b} The appropriate optimal waveform depends on the location of $\Omega$ with respect to $(\w_1,\w_2)$.  If $\Omega$ is in the Case I region then $v_-$ (resp. $v_+$) is optimal.  Otherwise, we use $v_*$, which depends on $\w_1$, $\w_2$ and $\Omega$.  Recall also the assumption $Q_*<0$. } \label{figarnold1}
\end{figure}

The boundaries of the Arnold tongues for $v_-$ and $v_+$ in (\ref{sol1}) are shown in Figure \ref{arnold3a}, and examples of oscillator ensembles for which they are optimal are indicated.   The optimal entrainment problem is fully characterized by the $PRC$ $Z$ and frequency range $(\w_1,\w_2)$ of $\cF$, as well as the target frequency $\Omega$.  To determine whether the problem is optimally solved by (\ref{sol1}), we derive the decision criterion by combining the definition $\Lambda_v(\vphi_+)=-\Delta\w_-$  with (\ref{bound1}) and (\ref{vminent1}) to obtain
\begin{equation} \label{omegrat1}
\frac{\Delta\w_-}{\Delta\w_+}=\frac{Q_*}{\bt{Z^2}}.
\end{equation}
This determines the boundary of the range of $\Omega$ in relation to $(\w_1,\w_2)$ when $v_-$ is optimal, and the derivation of the optimal range when $v_+$ is optimal is symmetric.  Therefore if
\begin{equation} \label{crit1}
\frac{\Delta\w_1}{\Delta\w_2} \geq \frac{Q_*}{\bt{Z^2}}  \quad \text{or} \quad \frac{\Delta\w_2}{\Delta\w_1} \geq \frac{Q_*}{\bt{Z^2}},
\end{equation}
then the minimum energy solution to problem (\ref{op2}) that entrains each member of $\cF$ is
\begin{equation} \label{sol2}
v(\theta)= \left\{\begin{array}{ll} \dS-\frac{\Delta \w_1}{\bt{Z^2}} Z(\theta+\vphi_+) &  \quad \text{if}\quad  \Delta\w_2<-\Delta\w_1, 
\\ \\\dS-\frac{\Delta \w_2}{\bt{Z^2}} Z(\theta+\vphi_-) &  \quad \text{if}\quad  -\Delta\w_1<\Delta\w_2, 
\end{array}\right.
\end{equation}

In practice we omit the phase ambiguity $\vphi_-$ or $\vphi_+$ in solution (\ref{sol2}).  The condition (\ref{crit1}) is illustrated as Case I in Figure \ref{arnold3b}.     When (\ref{crit1}) does not hold, then solution (\ref{sol2}) is not optimal for problem (\ref{op2}), as in example (C) in Figure \ref{arnold3a},which motivates the derivation of the optimal solution (\ref{sol5}) below.


\subsubsection{Case II: A difference of shifted PRCs is optimal for entrainment of the ensemble.}

To solve the (\ref{op2}) when (\ref{sol2}) is not optimal, as in (D) in Figure \ref{arnold3a}, we adjoin the constraints in (\ref{op2}) to the minimum energy objective function using multipliers $\mu_-$ and $\mu_+$, giving rise to the cost functional\\
\begin{eqnarray} \label{op3}
\cJ[v] & = & \bt{v^2}-\mu_-(\Delta \w_2+\Lambda_v(\vphi_-)) -\mu_+(-\Delta \w_1-\Lambda_v(\vphi_+)) \notag\\
& = & \bt{v^2}-\mu_-(\Delta \w_2+\bt{Z(\theta+\vphi_-)v(\theta)}) \notag\\ & & \qquad -\mu_+(-\Delta \w_1-\bt{Z(\theta+\vphi_+)v(\theta)}) \notag\\
& = & \frac{1}{2\pi}\int_0^{2\pi} \Bigl(v(\theta)[v(\theta) - \mu_-Z(\theta+\vphi_-) +\mu_+Z(\theta+\vphi_+)] \notag \\
& & \qquad -\mu_-\Delta\w_2+\mu_+\Delta\w_1 \Bigr)\rd \theta.
\end{eqnarray}\\
Solving the Euler-Lagrange equation yields
\begin{equation} \label{sol3}
v(\theta)=\dS-\frac{1}{2}[ \mu_+Z(\theta+\vphi_+)-\mu_-Z(\theta+\vphi_-)],
\end{equation}
which we substitute back into problem (\ref{op2}) to obtain
\begin{eqnarray} \label{eq3}
\bt{v^2}&=&\frac{1}{4}\langle\bp{\mu_+Z(\theta+\vphi_+)-\mu_-Z(\theta+\vphi_-)}^2\rangle  \\ & = & \frac{1}{4}\mu_+^2 \bt{Z^2} -\half\mu_+\mu_-\bt{Z(\theta+\vphi_+)Z(\theta+\vphi_-)} + \frac{1}{4}\mu_-^2 \bt{Z^2} \notag \\
& = & \frac{1}{4}(\mu_+^2+\mu_-^2)\bt{Z^2} - \half \mu_+\mu_-Q(\Delta\vphi), \notag\\\notag\\
\Lambda_v(\vphi_+) & = & \bt{Z(\theta+\vphi_+)v(\theta)} = -\half\mu_+\bt{Z^2} + \half\mu_-Q(\Delta\vphi), \label{eq4} \\
\Lambda_v(\vphi_-) & = & \bt{Z(\theta+\vphi_-)v(\theta)} = \half \mu_-\bt{Z^2}-\half\mu_+Q(\Delta\vphi). \label{eq5}
\end{eqnarray}\\
Substituting (\ref{eq3}), (\ref{eq4}), and (\ref{eq5}) into problem (\ref{op2}) simplifies the functional optimization problem to a nonlinear programming problem in the variables $\mu_-$, $\mu_+$, and $Q(\Delta\vphi)$, namely
\begin{equation} \label{op4}
\begin{array}{rr}
\min & \cJ[\mu_-,\mu_+,Q(\Delta\vphi)] = \dS\tfrac{1}{4}(\mu_+^2+\mu_-^2)\bt{Z^2} - \tfrac{1}{2} \mu_+\mu_-Q(\Delta\vphi) \medskip \\
\st &\dS\Delta \w_2+\tfrac{1}{2} \mu_-\bt{Z^2}-\tfrac{1}{2}\mu_+Q(\Delta\vphi)\leq 0, \medskip \\
 & \dS-\Delta \w_1+\tfrac{1}{2}\mu_+\bt{Z^2} - \tfrac{1}{2}\mu_-Q(\Delta\vphi)\leq 0.
\end{array}
\end{equation}
When one of the constraints is not active, then either $\mu_+=0$ (resp. $\mu_-=0$), and problem (\ref{op3}) is reduced to problem (\ref{op1}) with $\lambda=\mu_-$ (resp. $\lambda=-\mu_+$).  This occurs when condition (\ref{crit1}) holds, and then solution (\ref{sol2}) is optimal.  In the case that condition (\ref{crit1}) does not hold, it follows that both constraints in problem (\ref{op4}) are active.  The multipliers can be solved for, yielding
\begin{equation} \label{mult1}
\begin{array}{rcl}
\mu_+ & =&\dS\frac{2(\Delta\w_1\bt{Z^2}-\Delta\w_2 Q(\Delta\vphi))}{(\bt{Z^2}-Q(\Delta\vphi))(\bt{Z^2}+Q(\Delta\vphi))}, \medskip \\  \mu_- & =&\dS\frac{2(\Delta\w_1Q(\Delta\vphi)-\Delta\w_2\bt{Z^2}) }{(\bt{Z^2}-Q(\Delta\vphi))(\bt{Z^2}+Q(\Delta\vphi))}.
\end{array}
\end{equation}
For these multipliers, the objective in problem (\ref{op4}) is reduced to function of $Q=Q(\Delta\vphi)$ given by
\begin{eqnarray} \label{cost1}
\cJ[Q ] &=& \dS\frac{(\Delta\w_1\bt{Z^2}-\Delta\w_2 Q )^2+ (\Delta\w_1Q -\Delta\w_2\bt{Z^2})^2}{(\bt{Z^2}-Q )^2(\bt{Z^2}+Q )^2} \bt{Z^2}  \notag\\ & & - \dS\frac{2(\Delta\w_1\bt{Z^2}-\Delta\w_2 Q )(\Delta\w_1Q -\Delta\w_2\bt{Z^2}) Q }{(\bt{Z^2}-Q )^2(\bt{Z^2}+Q )^2}.
\end{eqnarray}
Differentiating the cost (\ref{cost1}) with respect to $Q$ results in\\
\begin{equation} \label{diff1}
\dxdy{\cJ[Q ]}{Q } = -2\frac{\Delta\w_1\Delta\w_2 Q ^2 - (\Delta\w_1^2+\Delta\w_2^2)\bt{Z^2}Q  + \Delta\w_1\Delta\w_2 \bt{Z^2}^2}{(\bt{Z^2}-Q )^2(\bt{Z^2}+Q )^2}.
\end{equation}
Because condition (\ref{crit1}) does not hold, it follows that $Q>\Delta\w_1\bt{Z^2}/\Delta\w_2$, and hence (\ref{diff1}) is positive for all admissible values of $Q(\Delta\vphi)$, so that the cost (\ref{cost1}) increases when $Q(\Delta\vphi)$ does.  Therefore the objective (\ref{cost1}) is minimized when $Q(\Delta\vphi)$ is, which occurs when $Q(\Delta\vphi)=Q_*=Q(\Delta\vphi_*)$.  Therefore the problem (\ref{op4}) is solved when $\Delta\vphi=\Delta\vphi_*$ and the multipliers are as in (\ref{mult1}). Therefore when
\begin{equation} \label{crit2}
\frac{\bt{Z^2}}{Q_*}\leq \frac{\Delta\w_1}{\Delta\w_2} \leq \frac{Q_*}{\bt{Z^2}},
\end{equation}
then Case II is in effect, and the minimum energy solution to problem (\ref{op2}) that entrains $\cF$ is \\
\begin{eqnarray} \label{sol4}
v(\theta)& = &\frac{(\Delta\w_2 Q_*-\Delta\w_1\bt{Z^2})}{(\bt{Z^2}-Q_*)(\bt{Z^2}+Q_*)}Z(\theta+\vphi_+) \notag\\ & & \quad+ \frac{(\Delta\w_1Q_*-\Delta\w_2\bt{Z^2})}{(\bt{Z^2}-Q_*)(\bt{Z^2}+Q_*)}Z(\theta+\vphi_-).
\end{eqnarray}\\
The locking range is exactly $R[k]=[\w_1,\w_2]$, satisfying the entrainment constraints (\ref{const2}).
In practice, we omit the phase ambiguity in the solution by using the equivalent control\\
\begin{eqnarray} \label{sol5}
v_*(\theta)& = &\frac{(\Delta\w_2 Q_*-\Delta\w_1\bt{Z^2})}{(\bt{Z^2}-Q_*)(\bt{Z^2}+Q_*)}Z(\theta+\Delta\vphi_*) \notag\\ & & \quad+ \frac{(\Delta\w_1Q_*-\Delta\w_2\bt{Z^2})}{(\bt{Z^2}-Q_*)(\bt{Z^2}+Q_*)}Z(\theta),
\end{eqnarray}\\
with energy
\begin{equation} \label{pow1}
\bt{v_*^2} = \frac{(\Delta\w_1^2+\Delta\w_2^2)\bt{Z^2}-2\Delta\w_1\Delta\w_2 Q_*}{(\bt{Z^2}-Q_*)(\bt{Z^2}+Q_*)}.
\end{equation}

\begin{figure}[t]
\centering \subfigure[]{\includegraphics[width=.6\linewidth]{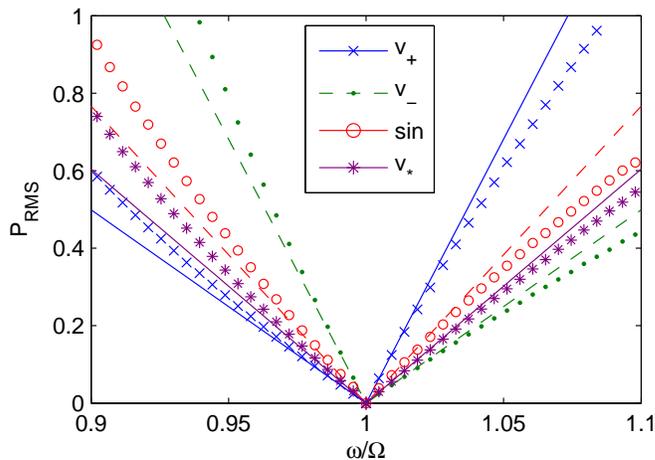} \label{figarnold2}} \\ \subfigure[]{\includegraphics[width=.6\linewidth]{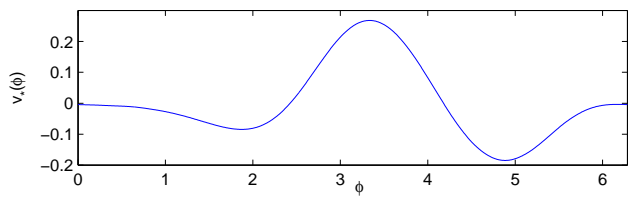} \label{voptc}} \caption{\ref{figarnold2}: Arnold tongues of Hodgkin-Huxley phase-model ensembles forced by $v_-$, $v_+$, a sine wave, and $v_*$ (shown in \ref{voptc}),  where the target $\Omega$ is the nominal natural frequency.  Lines are theoretical values and points are computed using a line search over the RMS forcing energy.  Note the difference with respect to Figure \ref{figatong1}.  Because the Arnold tongue characterizes entrainment of an ensemble of oscillators to a fixed target $\Omega$ with re-scaled natural frequency $\w/\Omega$ varying along the $x$-axis, its contours ``drift'' right as $P_{\rM{RMS}}$ increasese instead of left, as when the Arnold tongue characterizes entrainment of a single oscillator with fixed natural frequency with re-scaled target frequency $\Omega/\w$ varying along the $x$-axis. }
\end{figure}
The Arnold tongue for entrainment using $v_*$ can be generated by applying the approach used for $v_-$ in (\ref{ensat1}).  The advantage of the appropriate optimal control over a sinusoidal or square wave forcing input is illustrated in Figures \ref{figarnold1} and \ref{figarnold2}.  In the special case where $\Omega=\half(\w_1+\w_2)$, then the solution (\ref{sol5}) reduces exactly to the control $v_*$ that maximizes the locking range $R[k]$ for a fixed control energy, which is achieved when $\Delta\w_+=\Delta\w_-$ \cite{harada10,zlotnik11}.  The theoretical contribution presented here provides a clarification of the symmetry properties of that particular case.  The Arnold tongues that characterize entrainment properties of an ensemble $\cF$ are computed and plotted for several controls in Figure \ref{figarnold2}, which clearly demonstrates that $v_*$ can entrain an ensemble $\cF$ using a lower RMS energy than that required by a sinusoidal waveform.

\section{Robustness of entrainment to parameter uncertainty} \label{secneur}

In this section, we provide the results of several numerical simulations that further justify our approach to the entrainment of neural ensembles. In particular, we examine the effect of parameter variation on the phase response curve and optimal entrainment control for an ensemble of neurons.  We also provide a visualization of the uncertainty in the entrainment properties of a neuron ensemble that arises due to such parameter variation.  This is done by computing an Arnold tongue distribution, in which the minimum RMS energy required for entrainment of the ensemble to a given target frequency $\Omega$ is a random variable with a probability density on the positive real line, instead of a single value, for each $\w\in(\w_1,\w_2)$.  We demonstrate that the optimal entrainment waveform is minimally sensitive to variation in underlying system parameters, that it is always superior to a generic waveform such as a sinusoid or square pulse train, and that its amplitude can be appropriately chosen to entrain the neuron with the most problematic parameter set in the ensemble.

\begin{figure}[t]
\centering \subfigure[]{\includegraphics[width=.6\linewidth]{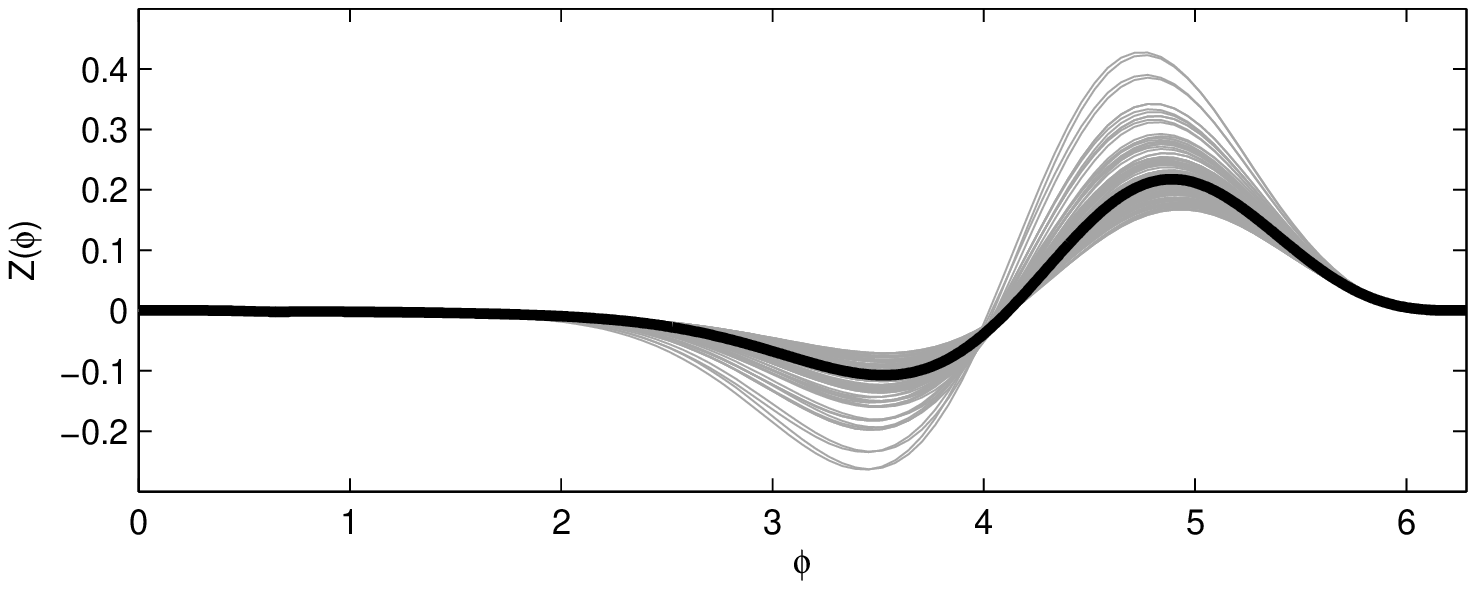} \label{crnprc1}} \\ \subfigure[]{\includegraphics[width=.6\linewidth]{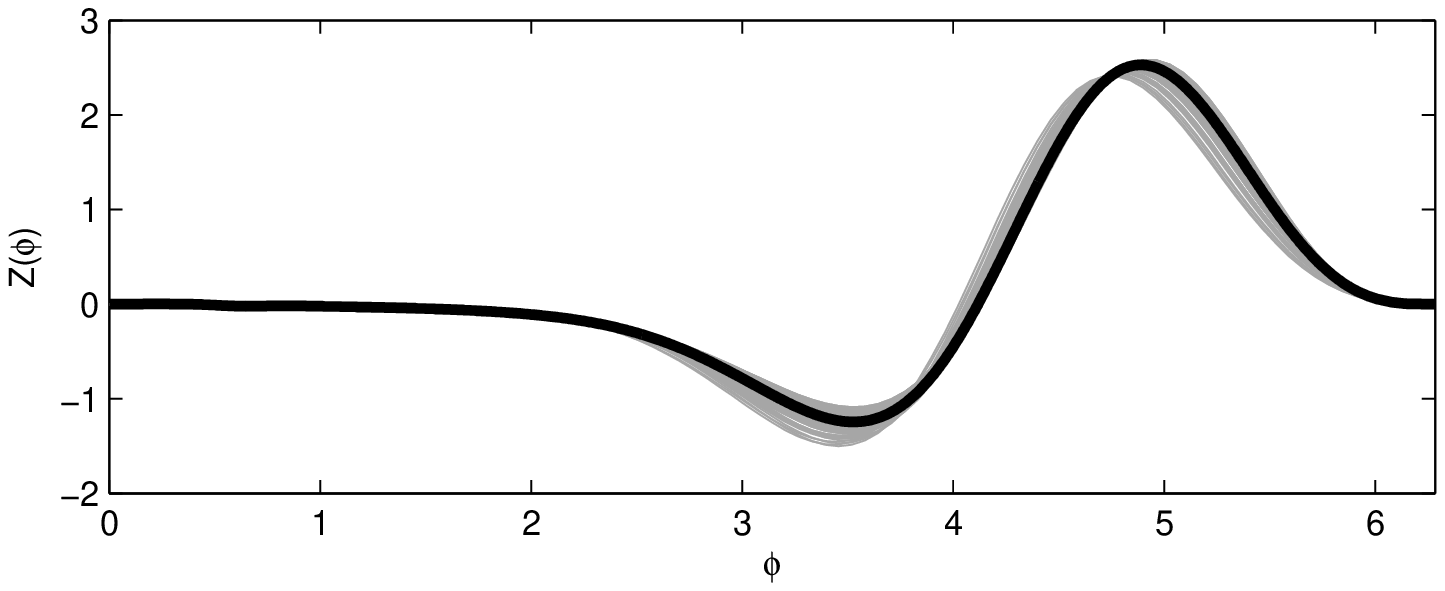} \label{crnprc2}} \caption{\ref{crnprc1} Phase response curves of the Hodgkin-Huxley neuron model and \ref{crnprc2} the same curves normalized to unity energy.  Each of the $2^7$ grey curves corresponds to a PRC obtained when each of the parameters $V_{Na}$, $V_K$, $V_L$, $\ol{g}_{Na}$, $\ol{g}_K$, $\ol{g}_L$, and $c$ is perturbed by 2\% above or below its nominal value.  In plot \ref{crnprc2} the PRCs of the perturbed systems are similar to the black curve, which is the PRC corresponding to the nominal parameter set.} \label{figcrnprc}
\end{figure}

The following sensitivity analysis can be performed to examine the effect of parameter variation on the PRC of an oscillator.  Suppose that $p_1,\ldots,p_d$ are the parameters that characterize the system dynamics, with nominal values $\alpha_1,\ldots,\alpha_d$.  We compute the PRC at each corner of a hypercube
$$
\cD = \prod_{i=1}^d (\beta_i,\gamma_i),
$$
where $(\beta_i,\gamma_i)$ is a small confidence interval for $\alpha_i$.  The corner points of $\cD$ are examined in particular in order to analyze the aggregate effect of uncertainty in all parameters.  If the $2^d$ curves so obtained are similar to the nominal PRC, then the optimal controls derived using the latter will be near optimal for entrainment of an uncertain ensemble.  Such a robustness property is important in practical neural entrainment applications, because biological oscillators exhibit significant variation from any nominal model.  Our analysis of the sensitivity of the Hodgkin-Huxley PRC to parameter variation appears in Figure \ref{figcrnprc}, in which we have used $(\beta_i,\gamma_i)=(0.98\alpha_i,1.02\alpha_i)$ for the $d=7$ parameter values in the model.  For each of the corner points of $\cD$ we plot the PRC, normalized to unity energy, and find that it does not vary significantly from the nominal curve.  This supports our assumption that the optimal entrainment waveform is minimally sensitive to variation in underlying system parameters.

\begin{figure}[t]
\centering \subfigure[]{\includegraphics[width=.6\linewidth]{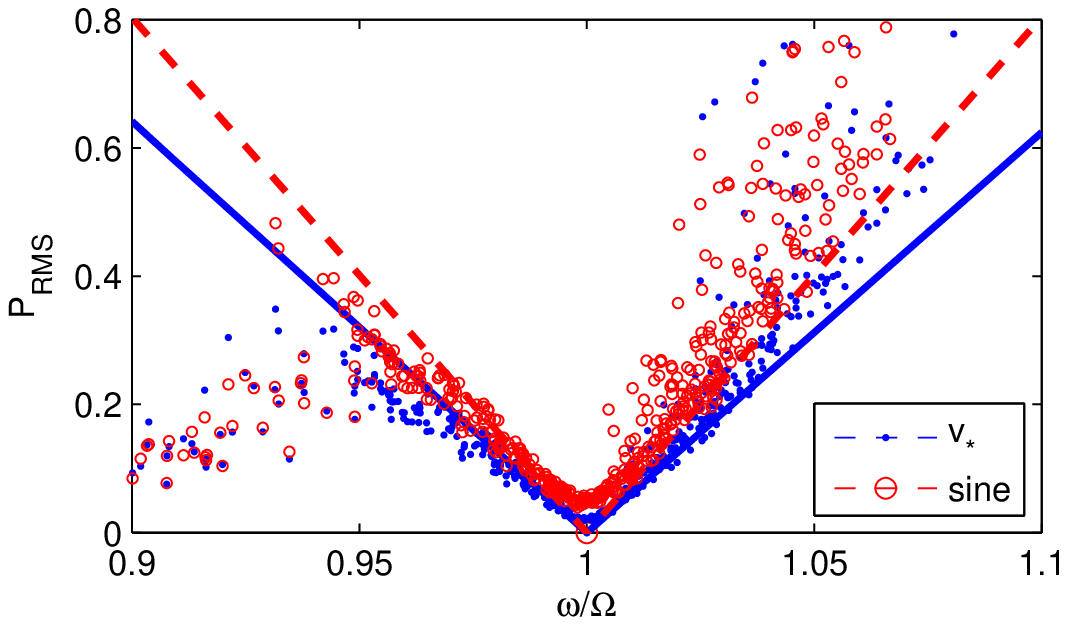} \label{earnold1}} \\ \subfigure[]{\includegraphics[width=.6\linewidth]{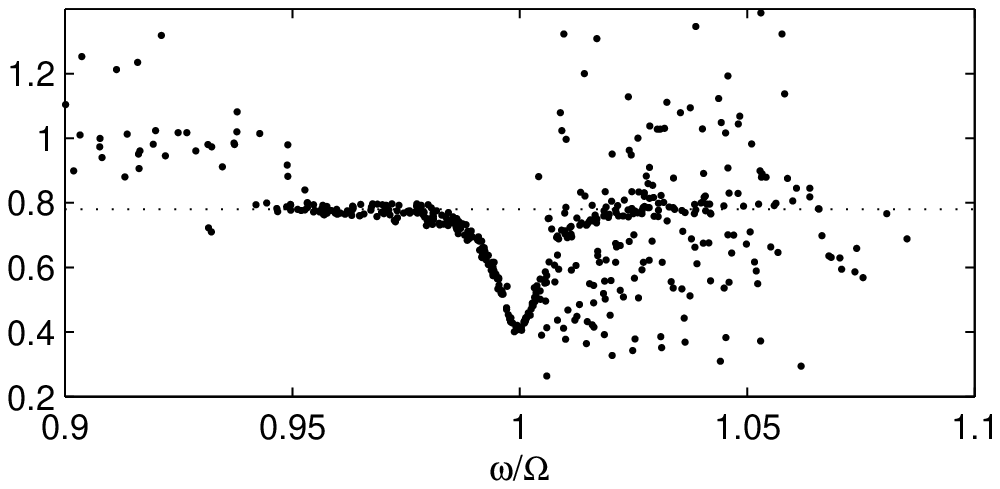} \label{earnold2}} \caption{\ref{earnold1} Arnold tongue distributions for an ensemble of Hodgkin-Huxley neurons with uniform random parameter variation on $(\beta_i,\gamma_i)=(.95\alpha_i,1.05\alpha_i)$, where $(\alpha_1,\ldots,\alpha_7)$ are nominal values for the parameter set $V_{Na}$, $V_K$, $V_L$, $\ol{g}_{Na}$, $\ol{g}_K$, $\ol{g}_L$, and $c$.  The target $\Omega$ is the nominal natural frequency.  There are $N=500$ randomly perturbed parameter sets used to generate empirical distributions (points) for entrainment with a sinusoid $v_s$ ($\circ$), and with $v_*$ ($\cdot$), the optimal waveform for entrainment to $\Omega=\half(\w_1+\w_2)$.  Solid lines are theoretical Arnold tongue boundaries.  \ref{earnold2} The ratio $P_k(v_*)/P_k(v_s)$ plotted as a function of $\w/\Omega$, the natural frequency of the neuron with parameter set $p_k$ (rescaled by the target frequency).  The optimal waveform requires an average of 22\% less RMS energy for entrainment than a sinusoid.}  \label{earnold}
\end{figure}

Although minor variations in Hodgkin-Huxley neuron model parameter values do not significantly effect the shape of the PRC, they do have a significant effect on the entrainment properties of a neuron ensemble by a fixed control waveform $v$.  The resulting uncertainty is visualized as an Arnold tongue distribution, which is the probability distribution of the minimum RMS control energy required to entrain an ensemble of oscillators, with parameter set distributed on a given probability space $\cD$, as a function of natural frequency $\w$.
In practice, we estimate this empirically for a hypercube $\cD$ with uniform probability measure by uniformly randomly generating samples $p_k\in \cD$ of the parameter for $k=1,\ldots,N$, for which we compute the natural frequency $\w_k$ of the perturbed Hodgkin-Huxley model and the minimum RMS control energy $P_k(v)$ required to entrain the $k\tH$ model using $v$.  This results in $N$ samples that are plotted on the energy-frequency plane, as shown in Figure \ref{earnold1}, which displays the empirical Arnold tongue distribution, using $N=500$ samples, for a sinusoidal control waveform and for the optimal waveform $v_*$ that maximizes the range of entrainment.  From visual inspection one concludes that the optimal waveform entrains the perturbed model using lower power than the sinusoid in the majority of cases.  More distinctly, Figure \ref{earnold2} displays the ratio between the minimum RMS energy levels required to entrain each parameter set $p_k$ using the optimal control and a sinusoid, as a function of the natural frequency $\w_k$.  Not only is the ratio below unity in most cases, with an average of 0.78, but there is also a clear trend line at 0.8 in the frequency range $0.95<\w<0.99$, with a much lower ratio for natural frequencies near the target $\Omega$.  This result strongly supports the assertion that the optimal ensemble entrainment waveform derived using our method is superior to traditional waveforms such as the sinusoid, not only for phase-reduced models, but also for the underlying non-reduced dynamical system model.

\begin{figure}[t]
\centerline { \includegraphics[width=.6\linewidth]{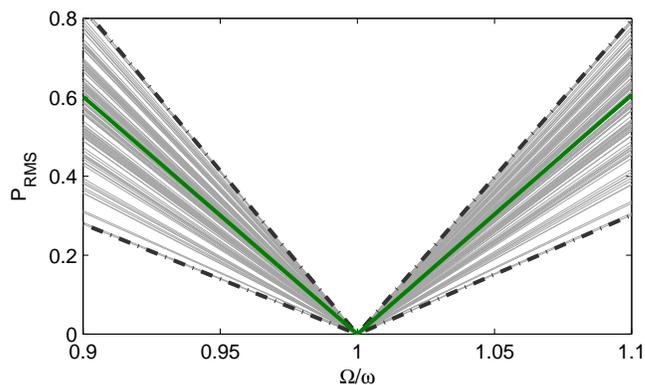}
} \caption{Theoretical Arnold tongues for entrainment of Hodgkin-Huxley neurons with parameter sets at the corner points of $\cD$ with $(\beta_i,\gamma_i)=(0.98\alpha_i,1.02\alpha_i)$, where the target $\Omega$ is the nominal natural frequency.  The Arnold tongues for the nominal parameter set (green) and the corner points (grey) are bounded by those of the best- and worst- case scenarios.  Each is generated using the PRC $Z_p$ for the parameter set $p$. }  \label{etarnold}
\end{figure}

In practice, given a confidence region hypercube $\cD$ for the parameter set $P$ of a neuron ensemble, the PRC can be computed at each corner point of $\cD$ and can be used to approximate the corresponding Arnold tongue when the perturbed system is entrained by the optimal waveform $v_*$ for the nominal parameter set.  In order to assure robust entrainment of the entire ensemble, the RMS energy of $v_*$ should be chosen such that it entrains the oscillator with the worst case parameter set, whose theoretical Arnold tongue is indicated by the top dashed line in Figure \ref{etarnold}.  The nominal oscillator is entrained with an RMS energy 24\% lower than the worst case scenario.  This difference is very near the average of 22\% less RMS energy that the optimal waveform requires to entrain an oscillator with 
parameter uncertainty.  It follows that simply by using the waveform $v_*$ instead of a square wave or sinusoid one can significantly enhance the likelihood that entrainment of an ensemble will be robust to such parameter variation.

\section{Conclusions} \label{secconc}

We have developed a method for minimum energy entrainment of oscillator ensembles to a desired frequency using weak periodic forcing.  Our approach is based on phase model reduction and formal averaging theory.  We also derive an approximation of the region of entrainability in energy-frequency space for an oscillator ensemble.  The entrainment of phase-reduced Hodgkin-Huxley neurons is considered as an example problem, and Arnold tongues are computed to evaluate the effectiveness of our controls.  The results closely match the theoretical bounds when the weak forcing requirement is fulfilled.  The optimal waveforms produce a similar result when applied to the original model, which suggests that optimal entrainment controls derived using a phase model are optimal for the original system, provided the oscillator remains within a neighborhood of its limit cycle.  We have justified our approach of considering an ensemble with common phase response curve and varying frequency by demonstrating robustness of the optimal waveforms to variation in model parameters.  The results of our simulations suggest that stimuli based on inherent dynamical properties of neural oscillators can result in significant improvement in energy efficiency and performance over traditional pulses in practical neural engineering applications.
This work furthermore provides a basis for evaluating the effectiveness of phase reduction techniques for the control of oscillating systems with parameter uncertainty.  In the future, we will extend the theory of optimal entrainment of oscillator ensembles to the case of $n:m$ entrainment, where $m$ cycles of the oscillator occur for every $n$ control cycles.  We will also derive controls that most rapidly entrain an ensemble of oscillators with uncertain initial state.  The approach described is of direct interest to researchers in chemistry and neuroscience, and may also be applied to vibration control in engineered systems.

\appendix

\section{Numerical Issues}

Because the PRC $Z(\theta)$ and forcing waveform $v(\theta)$ are both $2\pi$-periodic, we represent them using Fourier series,
\begin{equation}
\dS Z(\theta)= \dS \half a_0 + \sum_{n=1}^\infty a_n\cos(n\theta) + \sum_{n=1}^\infty b_n\sin(n\theta).
\end{equation}
\begin{equation}
\dS v(\theta)= \dS \half c_0 + \sum_{n=1}^\infty c_n\cos(n\theta) + \sum_{n=1}^\infty d_n\sin(n\theta).
\end{equation}

Using trigonometric angle sum identities and the orthogonality of the Fourier basis, we can express the interaction function $\Lambda_v$ as
\begin{align} \label{applamb}
\Lambda_v(\vphi)= \bt{Z(\theta+\vphi)v(\theta)} & = \frac{a_0c_0}{4}  + \half \sum_{n=1}^\infty [a_nc_n+b_nd_n]\cos(n\vphi) \nonumber \\ & \qquad +\half \sum_{n=1}^\infty [b_nc_n-a_nd_n]\sin(n\vphi).
\end{align}
In addition, for any $\vphi_1,\vphi_2\in[0,2\pi)$, periodicity of $Z$ and $v$ ensures that $\bt{Z(\theta+\vphi_1)v(\theta+\vphi_2)}=\bt{Z(\theta+\vphi_1-\vphi_2)v(\theta)}$.  Let $y=\cos(\vphi)$, so that we can write
\begin{align} \label{applambmod}
\Lambda_v(\vphi)= \ell_v(y) & = \frac{a_0c_0}{4}  + \half \sum_{n=1}^\infty [a_nc_n+b_nd_n]T_n(y) \nonumber \\ & \qquad +\half \sum_{n=1}^\infty [b_nc_n-a_nd_n]\sqrt{1-y^2}U_n(y).
\end{align}
where $T_n$ is the $n\tH$ Chebyshev polynomial of the first kind, and $U_n$ is the $n\tH$ Chebyshev polynomial of the second kind.  If $y_*$ is a minimizer of $\ell_v(y)$ over $y\in[-1,1]$, then $\vphi_*=\arccos(y_*)$ is a minimizer of $\Lambda_v(\vphi)$ over $\vphi\in[-\pi,\pi]$.  Recall that $Q(\Delta\vphi)=\bt{Z(\theta+\Delta\vphi)Z(\theta)}$ satisfies $Q(\Delta\vphi)=Q(-\Delta\vphi)$, so that
\begin{equation} \label{qexp}
Q(\Delta\vphi)= \frac{a_0^2}{4}  + \half \sum_{n=1}^\infty [a_n^2+b_n^2]\cos(n\Delta\vphi).
\end{equation}
To find $Q_*$ and $\Delta\vphi_*$ it is sufficient to minimize a truncation of the series in (\ref{qexp}) in the most numerically expedient way, whether explicitly as $Q(\Delta\vphi)=\Lambda_Z(\Delta\vphi)$ as in (\ref{applamb}) or by using the change of variables $y=\cos(\Delta\vphi)$ and minimizing $\ell_Z(y)$ as in (\ref{applambmod}), in which case the objective function is a polynomial with compact domain $[-1,1]\subset\bR$.

\section{Hodgkin-Huxley Model}

The Hodgkin-Huxley model describes the propagation of action potentials in neurons, specifically the squid giant axon, and is used as a canonical example of neural oscillator dynamics. The equations are
\begin{equation} \label{hheq}
\hskip-2pt\begin{array}{c}\begin{array}{rcl}
c\dot{V}&=&I_b+I(t)-\ol{g}_{Na}h(V-V_{Na})m^3-\ol{g}_K(V-V_k)n^4-\ol{g}_L(V-V_L)\\
\dot{m}&=&a_m(V)(1-m)-b_m(V)m,\\
\dot{h}&=&a_h(V)(1-h)-b_h(V)h,\\
\dot{n}&=&a_n(V)(1-n)-b_n(V)n,
\end{array}\\\\\begin{array}{rcl}
a_m(V)&=&0.1(V+40)/(1-\exp(-(V+40)/10)),\\ b_m(V)&=&4\exp(-(V+65)/18),\\
a_h(V)&=&0.07\exp(-(V+65)/20),\\ b_h(V)&=&1/(1+\exp(-(V+35)/10)),\\
a_n(V)&=&0.01(V+55)/(1-\exp(-(V+55)/10)),\\ b_n(V)&=&0.125\exp(-(V+65)/80).\\
\end{array}\end{array}
\end{equation}
The variable $V$ is the voltage across the axon membrane, $m$, $h$, and $n$ are the ion gating variables, $I_b$ is a baseline current that induces the oscillation, and $I(t)$ is the control input.  The units of $V$ are millivolts and the units of time are milliseconds. Nominal parameters are $V_{Na} =50 \text{ mV}$, $V_K=-77 \text{ mV}$, $V_L=-54.4 \text{ mV}$, $\ol{g}_{Na}=120 \text{ mS/cm}^2$, $\ol{g}_K = 36 \text{ mS/cm}^2$, $\ol{g}_L=0.3 \text{ mS/cm}^2$, $I_b=10 \,\,\mu\text{A/cm}^2$, and $c=1 \,\,\mu\text{F/cm}^2$, for which the period of oscillation is $T=14.63842\pm10^{-5}$ ms.

\section*{References}

\bibliographystyle{unsrt}
\bibliography{prc_bib}

\begin{thebibliography}{10}

\bibitem{strogatz01}
S.~Strogatz.
\newblock {\em Nonlinear Dynamics And Chaos: With Applications To Physics,
  Biology, Chemistry, And Engineering}.
\newblock Studies in nonlinearity. Westview Press, 1 edition, 2001.

\bibitem{hoppensteadt97}
F.~Hoppensteadt and E.~Izhikevich.
\newblock {\em Weakly connected neural networks}.
\newblock Springer-Verlag, New Jersey, 1997.

\bibitem{hanson78}
F.~Hanson.
\newblock Comparative studies of firefly pacemakers.
\newblock {\em Federation proceedings}, 38(8):2158--2164, 1978.

\bibitem{mirollo90}
R.~Mirollo and S.~Strogatz.
\newblock Synchronization of pulse-coupled biological oscillators.
\newblock {\em SIAM Journal on Applied Mathematics}, 50(6):1645--1662, 1990.

\bibitem{ermentrout84}
G.~Ermentrout and J.~Rinzel.
\newblock Beyond a pacemaker's entrainment limit: phase walk-through.
\newblock {\em American Journal of Physiology - Regulatory, Integrative and
  Comparative Physiology}, 246(1), 1984.

\bibitem{fischer00}
I.~Fischer, Y.~Liu, and P.~Davis.
\newblock Synchronization of chaotic semiconductor laser dynamics on
  subnanosecond time scales and its potential for chaos communication.
\newblock {\em Physical Review A}, 62, 2000.

\bibitem{blekhman88}
I.~Blekhman.
\newblock {\em Synchronization in science and technology}.
\newblock ASME Press translations, New York, 1988.

\bibitem{aronson86}
D.~Aronson, R.~McGehee, I.~Kevrekidis, and R.~Aris.
\newblock Entrainment regions for periodically forced oscillators.
\newblock {\em Physical Review A}, 33(3):2190--2192, 1986.

\bibitem{zalalutdinov03}
M.~Zalalutdinov, K.~Aubin, A.~Zehnder, R.~Hand, H.~Craighead, J.~Parpia, and
  B.~Houston.
\newblock Frequency entrainment for micromechanical oscillator.
\newblock {\em Applied Physics Letters}, 83(16):3281--3283, 2003.

\bibitem{berke04}
J.~D. Berke, M.~Okatan, J.~Skurski, and H.~B. Eichenbaum.
\newblock Oscillatory entrainment of striatal neurons in freely moving rats.
\newblock {\em Neuron}, 43:883--896, 2004.

\bibitem{sirota08}
A.~Sirota, S.~Montgomery, S.~Fujisawa, Y.~Isomura, M.~Zugaro, and G.~Buzs\'aki.
\newblock Entrainment of neocortical neurons and gamma oscillations by the
  hippocampal theta rhythm.
\newblock {\em Neuron}, 60:683--697, 2008.

\bibitem{tass03}
P.~A. Tass.
\newblock A model of desynchronizing deep brain stimulation with a
  demand-controlled coordinated reset of neural subpopulations.
\newblock {\em Biol. Cybern}, 89:81--88, 2003.

\bibitem{kuncel04}
A.~M. Kuncel and W.~M. Grill.
\newblock Selection of stimulus parameters for deep brain stimulation.
\newblock {\em Clinical Neurophysiology}, 115:2431--2441, 2004.

\bibitem{izhikevich06}
E.~Izhikevich and Y.~Kuramoto.
\newblock Weakly coupled oscillators.
\newblock In {\em Encyclopedia of mathematical physics}. Elsevier, 2006.

\bibitem{izhikevich07}
E.~Izhikevich.
\newblock {\em Dynamical Systems in Neuroscience}.
\newblock Neuroscience. MIT Press, 2007.

\bibitem{malkin49}
I.~Malkin.
\newblock {\em Methods of Poincare and Liapunov in the theory of nonlinear
  oscillations}.
\newblock Gostexizdat, Moscow, 1949.

\bibitem{kornfeld82}
I.~Kornfeld, S.~Fomin, and Y.~Sinai.
\newblock {\em Ergodic theory: Differentiable Dynamical Systems}, volume 245 of
  {\em Grund. Math. Wissens.}
\newblock Springer-Verlag, 1982.

\bibitem{kuramoto84}
Y.~Kuramoto.
\newblock {\em Chemical Oscillations, Waves, and Turbulence}.
\newblock Springer, New York, 1984.

\bibitem{pikovsky01}
A.~Pikovsky, M.~Rosenblum, and J.~Kurths.
\newblock {\em Synchronization: A Universal Concept in Nonlinear Science}.
\newblock Cambridge University Press, 2001.

\bibitem{strogatz00}
S.~H. Strogatz.
\newblock From kuramoto to crawford: exploring the onset of synchronization in
  populations of coupled oscillators.
\newblock {\em Physica D}, 143(1-4):1--20, 2000.

\bibitem{rosenblum96}
M.~G. Rosenblum, A.~S. Pikovsky, and J.~Kurths.
\newblock Phase synchronization of chaotic oscillators.
\newblock {\em Physical Review Letters}, 76(11):1804--1807, 1996.

\bibitem{hong02}
H.~Hong and M.~Y. Choi.
\newblock Synchronization on small-world networks.
\newblock {\em Physical Review E}, 65:026139(5), 2002.

\bibitem{pinsker77}
H.~M. Pinsker.
\newblock Aplysia bursting neurons as endogenous oscillators. i. phase-response
  curves for pulsed inhibitory synaptic input.
\newblock {\em J. Neurophysiology}, 40(3).

\bibitem{ermentrout96}
B.~Ermentrout.
\newblock Type i membranes, phase resetting curves, and synchrony.
\newblock {\em Neural Computation}, 8(5):979--1001, 1996.

\bibitem{perlmutter06}
J.~S. Perlmutter and J.~W. Wink.
\newblock Deep brain stimulation.
\newblock {\em Annu. Rev. Neurosci.}, 29:229--257, 2006.

\bibitem{good09}
L.~Good.
\newblock Control of synchronization of brain dynamics leads to control of
  epileptic seizures in rodents.
\newblock {\em International Journal of Neural Systems}, 19(3):173--196, 2009.

\bibitem{kiss02}
I.~Kiss, I.~Zhai, and J.~Hudson.
\newblock Emerging coherence in a population of chemical oscillators.
\newblock {\em Science}, 296:1676--1678, 2002.

\bibitem{nakata09}
S.~Nakata, K.~Miyazaki, S.~Izuhara, H.~Yamaoka, and D.~Tanaka.
\newblock Arnold tongue of electrochemical nonlinear oscillators.
\newblock {\em Journal of Physical Chemistry A}, 113:6876--6879, 2009.

\bibitem{hoppensteadt99}
F.~Hoppensteadt and E.~Izhikevich.
\newblock Oscillatory neurocomputers with dynamic connectivity.
\newblock {\em Physical Review Letters}, 82(14), 1999.

\bibitem{hunter03}
J.~Hunter and J.~Milton.
\newblock Amplitude and frequency dependence of spike timing: Implications for
  dynamic regulation.
\newblock {\em Journal of Neurophysiology}, 90:387--394, 2003.

\bibitem{ullah09}
G.~Ullah.
\newblock Tracking and control of neuronal hodgkin-huxley dynamics.
\newblock {\em Physical Review E}, 79:040901(R), 2009.

\bibitem{nabi09}
A.~Nabi and J.~Moehlis.
\newblock Charge-balanced optimal inputs for phase models of spiking neurons.
\newblock In {\em 2009 ASME Dynamic Systems and Control Conference}, Hollywood,
  CA, October 2009.

\bibitem{dasanayake11}
I.~Dasanayake and J.-S. Li.
\newblock Optimal design of minimum-power stimuli for phase models of neuron
  oscillators.
\newblock {\em Physical Review E}, 83:061916, 2011.

\bibitem{dasanayake12}
I.~Dasanayake and J.-S. Li.
\newblock Charge-balanced minimum-power controls for spiking neuron
  oscillators.
\newblock {\em IEEE Transactions on Automatic Control (under review)}.

\bibitem{hodgkin52}
A.~Hodgkin and A.~Huxley.
\newblock A quantitative description of membrane current and its application to
  conduction and excitation in nerve.
\newblock {\em The Journal of Physiology}, 117(4), 1952.

\bibitem{li06thesis}
J.-S. Li.
\newblock {\em Control of Inhomogeneous Ensembles}.
\newblock PhD thesis, Harvard University, Cambridge, MA, 2006.

\bibitem{harada10}
T.~Harada, H.~Tanaka, M.~Hankins, and I.~Kiss.
\newblock Optimal waveform for the entrainment of a weakly forced oscillator.
\newblock {\em Physical Review Letters}, 105(8), 2010.

\bibitem{zlotnik11}
A.~Zlotnik and J.~Li.
\newblock Optimal asymptotic entrainment of phase-reduced oscillators.
\newblock In {\em 2011 ASME Dynamic Systems and Control Conference}, Arlington,
  VA, October 2011.

\bibitem{brown04}
E.~Brown, J.~Moehlis, and P.~Holmes.
\newblock On the phase reduction and response dynamics of neural oscillator
  populations.
\newblock {\em Neural Computation}, 16(4):673--715, 2004.

\bibitem{efimov10}
D.~Efimov and T.~Raissi.
\newblock Phase resetting control based on direct phase response curve.
\newblock In {\em Preprints of the 8th IFAC Symposium on Nonlinear Control
  Systems}, pages 332--337, Bologna, September 2010.

\bibitem{efimov09}
D.~Efimov, P.~Sacr\'e, and R.~Sepulchre.
\newblock Controlling the phase of an oscillator: A phase response curve
  approach.
\newblock In {\em Joint 48th Conference on Decision and Control}, pages
  7692--7697, December 2009.

\bibitem{perko90}
L.~Perko.
\newblock {\em Differential equations and dynamical systems}.
\newblock Texts in applied mathematics. Springer, 2 edition, 1990.

\bibitem{kelley04}
W.~Kelley and A.~Peterson.
\newblock {\em The Theory of Differential Equations, Classical and
  Qualitative}.
\newblock Pearson, 2004.

\bibitem{aprille72}
T.~Aprille and T.~Trick.
\newblock A computer algorithm to determine the steady-state response of
  nonlinear oscillators.
\newblock {\em IEEE Transactions on Circuit Theory}, 19(4):354--360, 1972.

\bibitem{khalil02}
H.~Khalil.
\newblock {\em Nonlinear Systems}.
\newblock Prentice Hall, 3 edition, 2002.

\bibitem{peressini00}
A.~Peressini, F.~Sullivan, and J.~Uhl.
\newblock {\em Mathematics of Nonlinear Programming}.
\newblock Springer, 2000.

\bibitem{govaerts06}
W.~Govaerts and B.~Sautois.
\newblock Computation of the phase response curve: A direct numerical approach.
\newblock {\em Neural Computation}, 18(4):817--847, 2006.

\bibitem{ermentrout02}
B.~Ermentrout.
\newblock {\em Simulating, Analyzing, and Animating Dynamical Systems: A Guide
  to XPPAUT for Researchers and Students}.
\newblock SIAM.

\bibitem{gelfand00}
I.~Gelfand and S.~Fomin.
\newblock {\em Calculus of Variations}.
\newblock Dover, 2000.

\end{thebibliography}

\end{document}